\documentclass[10pt,twocolumn,a4paper]{article}

\usepackage[left=15mm, right=15mm, top=15mm, bottom=15mm]{geometry}
\usepackage{algorithm,algcompatible,amsmath,amssymb, amsthm}

\usepackage{subfig}
\usepackage{flushend}
\usepackage{indentfirst}
\usepackage{graphics}
\usepackage{amsmath}
\usepackage{graphicx}
\usepackage{epstopdf}
\usepackage{float}
\usepackage{listings}

\usepackage[font=footnotesize,labelfont=bf]{caption}
\usepackage[labelsep=period]{caption}
\theoremstyle{definition}
\newtheorem{definition}{Definition}[section]
\theoremstyle{theorem}
\newtheorem{theorem}{Theorem}[section]

\graphicspath{{./figure/}}

\begin{document}

\title{\huge \textbf{Theoretical Justification of the Bi Error Method}}

\twocolumn[
\begin{@twocolumnfalse}

\author{\textbf{MaryLena Bleile}$^{1, *}$\\\\
\footnotesize $^{1}${Department of Statistical Science, Southern Methodist University, Dallas, 75205, Texas, United States}\\
\footnotesize $^{*}$Corresponding Author: mbleile@smu.edu}

\date{}

\maketitle

\begin{flushleft}
\noindent \footnotesize {Copyright \copyright 2020 by authors, all rights reserved. Authors agree that this article remains permanently\\
\noindent \footnotesize open access under the terms of the Creative Commons Attribution License 4.0 International License}
\end{flushleft}

\end{@twocolumnfalse}
]

\noindent \textbf{\large{Abstract}} \hspace{2pt}Incorrect usage of $p$-values, particularly within the context of significance testing using the arbitrary .05 threshold, has become a major problem in modern statistical practice. The prevalence of this problem can be traced back to the context-free 5-step method commonly taught to undergraduates: we teach it because it is what is done, and we do it because it is what we are taught. This hold particularly true for practitioners of statistics who are not formal statisticians. Thus, in order to improve scientific practice and overcome statistical dichotomania, an accessible replacement for the 5-step method is warranted. We propose a method foundational on the utilization of  the Youden Index as a potential decision threshold, which has been shown in the literature to be effective in conjunction with neutral zones. Unlike the traditional 5-step method, our 5-step method (the Bi Error method) allows for neutral results, does not require $p$-values, and does not provide any default threshold values. Instead, our method explicitly requires contextual error analysis as well as quantification of statistical power. Furthermore, and in part due to its lack of usage of p-values, the method sports improved accessibility. This accessibility is supported by a generalized analytical derivation of the Youden Index.
\\

\noindent \textbf{\large{Keywords}} \hspace{2pt} Youden Index, p-value, Type II Error\\

\noindent\hrulefill

\section{\Large{Introduction}}

The problematic nature of blindly executing hypothesis tests using the $p<$ .05 has been clearly illustrated: the  problem received official recognition by the ASA in 2016, culminating in the recent release of a special edition of The American Statistician, which featured 43 papers on the topic \cite{ASAstatement, AmStatissue} . Subsequently, the National Institute of Statistical Science hosted a webinar on alternatives to the p-value, which was followed up in November with a more in-depth discussion featuring three experts in the field: Jim Berger, Sander Greenland, and Robert Matthews \cite{webinarMay, webinarFall}. Cobb points out the cyclic nature of the problem: ``We teach it because it’s what we do, we do it because it’s what we teach” \cite{ASAstatement}.\par 
Yet in spite of the extensive literature on the adverse effects of procedural, context-blind statistical hypothesis testing with $p<$.05, we continue to teach (and sometimes practice) the 5-step Null-Hypothesis Significance Testing (NHST) method outlined in Algorithm \ref{NHST}.

\begin{algorithm}
\caption{NHST Method}
\label{NHST}
  \begin{algorithmic}[1]
    \STATE Specify the null hypothesis
    \STATE Specify the alternative hypothesis
    \STATE Set the significance level (usually $\alpha = .05$)
    \STATE Calculate the test statistic and the corresponding p-value
    \STATE If the p-value is less than the significance level, then reject the null hypothesis and conclude statistical significance. Otherwise, fail to reject the null hypothesis and conclude statistical non-significance.
  \end{algorithmic}
\end{algorithm}
Of course, the field of statistical methodology is rich and extensive beyond this. Specifically, hypothesis testing with neutral zones has shown to be effective. The idea behind this strategy is to allow for a third option aside from rejecting or failing to reject the null hypothesis $H_0$: \textit{inconclusive results}. However, many of the methods from the statistical literature lack the simplicity and algorithmic structure of the existing 5-step method, which is perhaps what contributes to its attractiveness for teaching and for use by non-statisticians. To quote Wasserstein, Schirm, \& Lazar, ```Don’t’ is not enough”: it is not enough to simply ban scientists and statisticians from using or teaching the existing 5-step method, leaving a gap in the curriculum. In order to fill this gap, we propose a replacement 5-step method for use in place of the above\cite{Beyondp}. \par
Statistical hypothesis tests can be considered as binary classifiers between the null and alternative hypotheses. The Receiver Operating Characteristic curve, which plots sensitivity vs 1-specificity is of special interest when discussing classifiers (here, sensitivity is the propensity of the test to reject the null given that the null is actually false, whereas specificity is the propensity of the test to \textit{fail} to reject the null, given that the null is actually true). Naturally, adding these two quantities together results in a value analogous to the chance of making a correct decision. Note, both sensitivity and specificity depend only on what decision threshold one uses to classify. The Youden Index is defined as $J = max_c(Sensitivity(c) + Specificity(c) - 1)$, where $c$ is the decision threshold \cite{youden}. That is, $J$ is the decision threshold which maximizes the aggregated sensitivity and specificity of the classifier. \par
\begin{figure}[thbp]
\centering
\includegraphics[width=4.6cm]{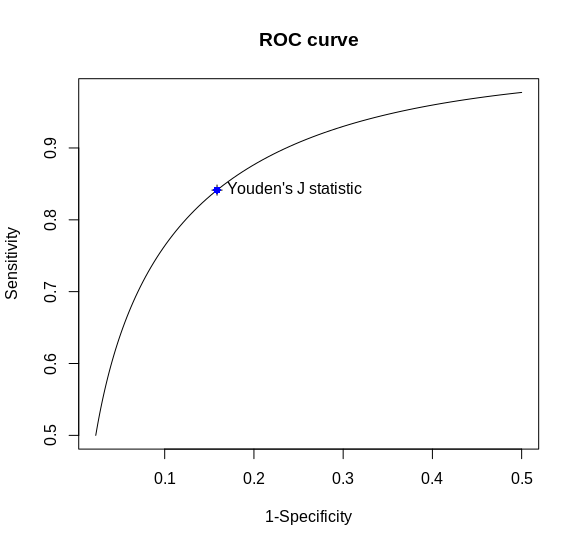}
\caption{An ROC curve illustrating the Youden Index.}
\label{ROC}
\end{figure}
In other words, $J$ is the decision threshold which minimizes the additive Type I and Type II error (where Type I error, denoted $\alpha$, is the probability of falsely rejecting the null, and Type II error, denoted  $\beta$, is the probability of failing to reject when the null is false). That is $J = minarg_c(\alpha+ \beta)$ (note, $\alpha, \beta$ are both functions of the decision threshold $c$). Youden's Index has been widely used in the context of hypothesis testing with neutral zones \cite{neutral}. \par
Algorithm \ref{BiError} outlines the proposed method. We call this method the ``Bi Error Method" for the two following reasons:
\begin{enumerate}
    \item Youden's $J$ minimizes two error terms
    \item Subjective analysis of two error rates is required
\end{enumerate}
\begin{algorithm}
\caption{The Bi Error Method}
\label{BiError}
  \begin{algorithmic}[1]
    \STATE Identify the null and alternative sampling distributions
    \STATE Set reasonable upper thresholds for Type I and Type II error ($\alpha, \beta$, respectively).
    \STATE Find (or approximate) the Youden Index ($J$)
    \STATE If the values of $\alpha, \beta,$ (and, implicitly, $\zeta (J)$)are reasonable, proceed to 5. Else, if any one of the above is too high, the results are inconclusive.
    \STATE If $\alpha, \beta,$ are reasonable and the observed test statistic is more extreme than $x_0$, then reject $H_0$. Else, fail to reject $H_0$.
  \end{algorithmic}
\end{algorithm}

We provide 5 of the most compelling reasons why the proposed method is attractive.\par
\underline{Accessibility.} Unlike some of the more sophisticated methodology used in the statistical literature, our method has the same 5-step, algorithmic structure as the existing method. \par
\underline{No p-values.} The pervasiveness and negative effects of misinterpretations of p-values in the academic literature is well-documented \cite{NotLearnt, errors, dichotomania, psychwrong, ASAstatement}. Our method eliminates the need for their use, which serves to potentially increase the accuracy of statistical practice in scientific research.\\
\underline{Contextual Error Analysis.} One of the ASA’s recommendations for hypothesis testing was that context should be considered in the analysis \cite{ASAstatement}. Yet, we still run into issues where the hypothesis testing is blindly performed with no attention to context, which has led to potentially disastrous results \cite{errors, dichotomania, covid}. Our method propagates ``Thoughtful research” as discussed in the literature, by including an explicit step involving contextual analysis, with no default thresholds for any of the values analyzed \cite{Beyondp}.\par 
\underline{Inconclusive results option.} The method propagates the acceptance of uncertainty in another way. In congruence with methods that have been shown to be effective in the literature, it explicitly allows for a third option besides rejecting or failing to reject the null hypothesis: \textit{inconclusivity} \cite{neutral, YoudeNeutral}.\par
\underline{Quantitative incorporation of Type II error}. One of the major issues with the statistical hypothesis testing method currently in place, is that it does not take into account the probability of falsely failing to reject: aka Type II error \cite{errors, covid}. In contrast, through the utilization of the Youden index, this method explicitly incorporates this quantitatively into the analysis.

\section{\Large{Materials and Methods}}
Since our method involves contextual, subjective error analysis, it is impossible to simulate its true performance. We can, however, get a general idea of what its lower bound on performance might be by simulating the hypothesis testing scenario and dichotomously classifying results based on whether they are larger than the Youden index.\par
It is notable that since we are utilizing the Type II error in the test procedure, hypothesis tests must be performed with a point alternative. This is desirable since it forces the researcher to think about effect size: here the alternative parameter does not necessarily represent what the population parameter must be if the test rejects the null hypothesis; rather, it represents such an effect that would be cared about within the context of the study. So, for the hypothesis test of:\\
\begin{center}
    $H_0: \theta = \theta_0$ vs.
    $H_A: \theta = \theta_A$
\end{center}
there are actually \textit{three} quantities of interest: $\theta$, $\theta_0,$ and $\theta_A$, where $\theta$ is the actual, unobserved value of the parameter of interest, and $\theta_0, \theta_A$ are the researcher-determined point null and alternative values, where $\theta_A$ is determined by the effect the researcher is expecting to see. The simulation was structured as follows:
\begin{algorithm}
    \caption{Monte Carlo Simulation}
  \begin{algorithmic}[1]
    \STATE Generate a sample of size $n$ from a normal distribution with mean $\mu$, variance 1 
    \STATE Perform a one-sample t-test of the hypothesis $H_0: \mu = 0$
    \STATE Check whether the mean is greater than $\hat{\mu_A}/2$, where $\hat{\mu_A}$ is the ``hypothesized" alternative mean. If so, reject $H_0$.
    \STATE Repeat for all combinations of values of $n, \mu,$ and $\hat{\mu_A}$
  \end{algorithmic}
\end{algorithm}

We tested for $n=10,20,30,50$, $\mu = 0, 0.3,0.5,0.7,1$, and $\hat{\mu_A} = 0.3,0.5,0.7,1,1.5,2,2.5,3,3.5$. The entire process was repeated a total of $M = 10 000$ times. Simulation was conducted using the statistical package R \cite{R}.\\

\section{\Large{Results}}
As expected, simulation results showed that the Bi Error method generally outperformed the traditional NHST method in terms of accuracy, particularly when the hypothesized alternative mean was greater than the actual mean. Tabular results and the simulation code are available in appendices A and B, respectively.\par 
In order to simplify step 3 for researchers and students, we also provide a generalized derivation of the Youden Index. Suppose we are testing:\\
 $H_0: \theta \leq \theta_0$ vs\\
 $H_A: \theta >\theta_0$,  where $\theta$ is an unknown parameter and $\theta_0$ is a value.\\ We restrict our attention to this case for simplicity, since the extension of our results to the case where $\theta < \theta_0$ and the two-tailed case is trivial.\\
 
\begin{theorem}
\label{main}
Suppose $f_0, f_A$ are the sampling distributions of a statistic $X$ under the null and alternative hypotheses, respectively, such that  $f_0,f_A$ are symmetric and satisfy $\vert \mu - x_1 \vert > \vert \mu - x_2 \vert \implies f(\vert \mu - x_1 \vert) < f(\vert \mu - x_2 \vert )$ (that is, they decrease in the tails).
Then $J = (\mu_0 + \mu_A)/2$.
\end{theorem}
This result has been previously shown for the Normal distribution, which is a special case \cite{YoudenMath, partialYouden, YoudeNormal}. However, Theorem \ref{main} is more general in that it applies to a wide range of distributions, including the case where the mean and/or variance do not exist. \\
Before presenting the proof, it is convenient to introduce the following definition:
\begin{definition}
The \textit{Bi Error} is defined as $\zeta(c) = \alpha + \beta$ 
\end{definition}

We now present the proof of Theorem \ref{main}.
\begin{proof}
Note, it suffices to minimize $2\zeta : = \zeta_2= 1 - F_0(x) + F_A(x) $. Assume, without loss of generality, that $\mu_A \geq \mu_0$. In this case, a critical value less than $\mu_0$ is not useful, so we will restrict our attention to $x>\mu_0$.  So, $\frac{\partial \zeta_2}{\partial x} = -f_0(x) + f_A(x)$. Setting this equal to zero yields $f_0(x) = f_A(x)$.\\
\begin{figure}[thbp]
\centering
\includegraphics[width=4.6cm]{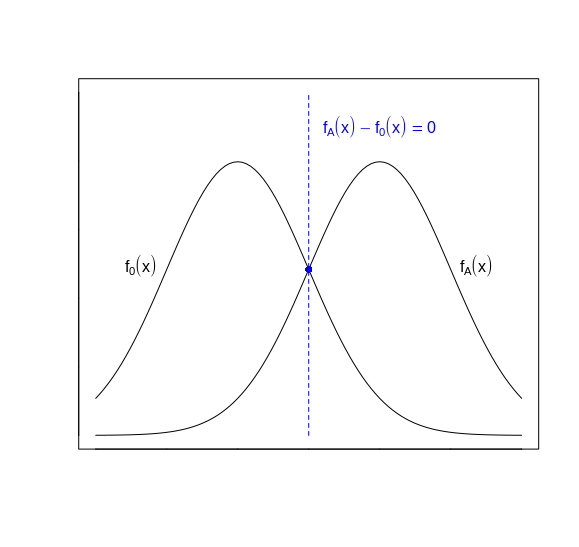}
\caption{The stationary point is found at $f_0(x) = f_A(x)$.}
\label{setup}
\end{figure}
Since $f_0, f_A$ are location family, this is equivalent to saying $f_0(x) = f_0(x-\delta)$, where $\delta = \mu_A - \mu_0$. But $f_0$ is unimodal. So the mode must be in the interval $(x-\delta, x)$. Also, since $f_0$ is symmetric, $\delta = 2(x - \mu_0$)\\
\begin{figure}[thbp]
\centering
\includegraphics[width=4.6cm]{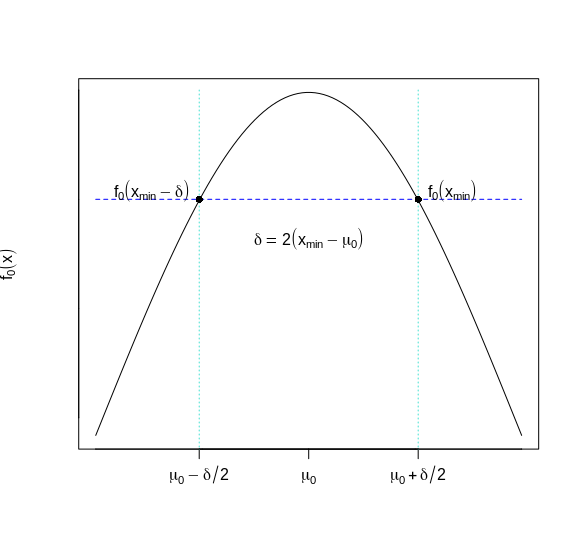}
\caption{Heuristic justification for $\delta = 2(x - \mu_0)$.}
\label{delta}
\end{figure}
$\implies \mu_A - \mu_0 =2(x-\mu_0)$\\
$\implies \frac{\mu_A + \mu_0}{2} = x$. \\
Since $f_0$ is univariate, we know that this stationary point must be a maximum or a minimum. In order to show it is a maximum, we will show that the derivative is negative in a range antecedent to the point $x_0 = \frac{\mu_A + \mu_0}{2}$, and positive in a range immediately before it. \par
Now we will show
\begin{equation}
    \forall \epsilon>0, \zeta_2'(\frac{\mu_0+\mu_A}{2}+\epsilon) > 0
\end{equation}
Consider $z'(x_0+\epsilon) = -f_0(\frac{3\mu_A-\mu_0}{2}+\epsilon) + f_A(\frac{3\mu_A-\mu_0}{2}+\epsilon)$. So, in order for (1) to hold, we need $\vert f_A(\frac{3\mu_A-\mu_0}{2}+\epsilon)\vert < \vert f_A(\frac{\mu_A+\mu_0}{2}+\epsilon)\vert$. But since $f_A$ is symmetric and decreases in the tails, this holds if $\frac{3\mu_A-\mu_0}{2}\geq \frac{\mu_0+\mu_A}{2} + 2(\mu_A - \frac{mu_A-\mu_0}{2})$. But the term on the right of that equation simplifies to $\frac{3\mu_A-\mu_0}{2}$ as required. An isomorphic argument exists for $f'(z-\epsilon) <0$.\\
\end{proof}

\section{\Large{Discussion}}
\subsection{\normalsize \textbf{On the Choice of Hypothesized $\delta$}}
One glaringly obvious possible criticism of this method lies in the necessary choice of ``alternative" distribution ($\hat{\mu_A}$, or equivalently $\delta = \mu_A - \mu_0$). Furthermore, it does not make sense to directly estimate this parameter from the data, (using it twice: once to estimate effect size, and then again to perform the test) since in the case of a symmetric distribution, if the null hypothesis is \textit{actually} true, this will cause us to use a critical value of approximately $(\mu_0+ \hat{\mu_A})/2 =  2\mu_0/2 = \mu_0$, which is obviously not useful, since it causes a Type I error of $\alpha=.5$. \par
In light of this, $\hat{\mu_A}$ should be, rather, chosen by the researcher in light of the context of the study (i.e. $\delta$ should be what effect, if one exists, that they are expecting). Furthermore, the estimate of $\mu_A$ should be identified \textit{before} the analyst explores the data. \par
Note, if $\hat{\mu_A}$ is too close to $\mu_0$, this threatens the experiment with inconclusive results due to an unreasonably high Type I error. Similarly, if $\hat{\mu_A}$ is too far from $\mu_0$, this threatens the experiment with inconclusive results due to an unreasonably high Type II error.\par
The simulated results indicate that the method performs particularly well when the researcher marginally \textit{overestimates} the expected effect, aka when $\mu_A > \mu$. We believe this lends additional credence to the validity of our method due to the fact that it appears to take advantage of a cognitive bias, but we defer a thorough discussion of this to a subsequent paper.

\subsection{\normalsize \textbf{Normal Distribution}}
\begin{figure}[thbp]
\centering
    \includegraphics[width=4.6cm]{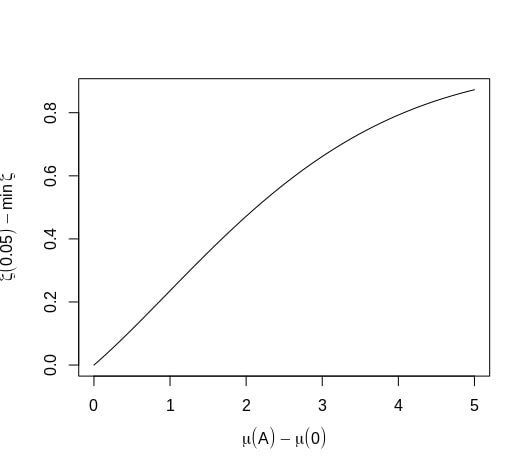}
  \caption{ Difference in $\zeta$ by $\mu_A - \mu_0$, with fixed $\sigma^2 = 2$}
  \label{fig:fixedsigma}
\end{figure}

\begin{figure}[thbp]
\centering
  \includegraphics[width=4.6cm]{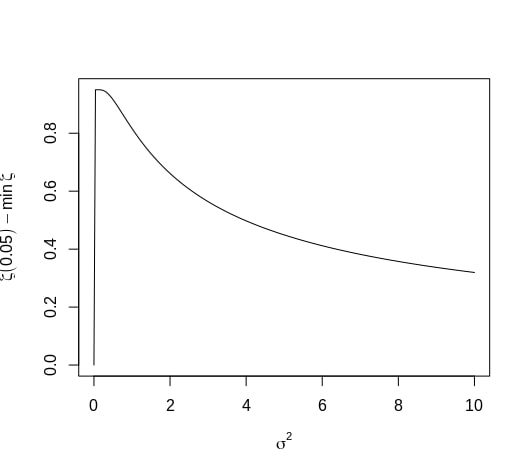}
  \caption{ Difference in $\zeta$ by $\sigma^2$, with fixed $\mu_A - \mu_0 = 3$}
  \label{fig:fixedmudiff}
\end{figure}

We have shown that the Bi Error is minimized when we choose $(\mu_0 + \mu_A)/2$ as a critical value. A natural question is, what difference does this make, as compared with conventional methods? That is, what does the Bi Error look like when we choose a critical value corresponding to $\alpha = 0.05$?\\
As suggested in the introduction, consider the case where $f_0,f_A$ are normal distributions, with different location parameters. Clearly the normal distribution satisfies the conditions necessary for Theorem 1 to hold. We are interested in the change  in Bi Error due to choosing  $\alpha, \beta$ such that $\zeta$ is minimized, as opposed to choosing $\alpha = 0.05$. More formally, we want to look at $\xi = \zeta(z_{0.05}) - \zeta((\mu_0+\mu_A)/2)$, where $z_0.05$ is the 95th percentile for $N(\mu_0, \sigma^2)$.\\
Plotting with fixed $\sigma^2$ revealed that $\xi$ has a positive relationship with $\delta = \mu_A - \mu_0$, as shown in Fig. \ref{fig:fixedsigma}. Plotting with fixed $\mu_A, \mu_0$ and letting $\sigma^2$ vary revealed an inverse relationship between $\xi$ and $\delta$ (Fig. \ref{fig:fixedmudiff}). So, by definition, $\xi$ increases with effect size.\\

\begin{figure}[thbp]
\centering
  \includegraphics[width=8.5cm]{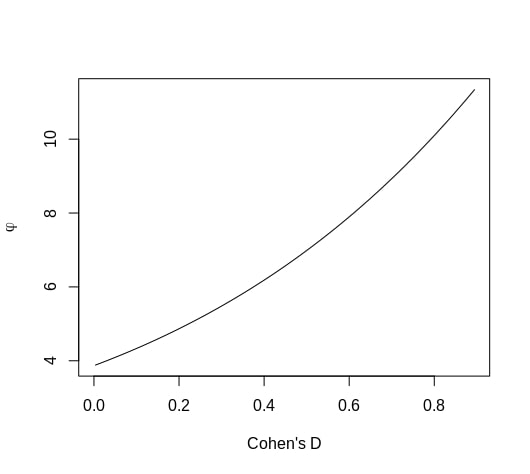}
  \caption{ Cohen's D vs $\phi$, with $\sigma^2=5, \mu_0=0$, and $\mu_A$ ranges from $0$ to $2$}
  \label{fig:dphi}
\end{figure}

Since $\zeta$ is a probability, it is bounded. Also, the raw difference may have different meanings in different scenarios, depending on the actual values of $\zeta$. So, it makes more sense to look at the odds ratio, $\omega1/\omega2$, where $\omega1 = \zeta(z_{0.05}/(1-\zeta(z_{0.05})), \omega2 = \zeta(0.5(\mu_0+\mu_A))/(1- \zeta(0.5(\mu_0+\mu_A)))$. Plotting $\phi_\zeta = \omega1/\omega2$ with effect size (Cohen's D), yields Fig. \ref{fig:dphi} wherein it is notable that, even for small effect sizes, the odds of making an incorrect conclusion on a hypothesis test are $4-6$ times larger if we set $\alpha=0.05$, rather than minimizing the Bi Error function.\\

This phenomenon is explained in greater depth when we consider the actual algebraic representation of $\xi$. Define $\zeta_\alpha$ to be the Bi Error for a standard $z$ test using the usual critical value $x_\alpha$, chosen to achieve some fixed Type I error $\alpha$. Similarly, define $\zeta*$ to be the Bi Error for the same scenario, but using Theorem \cite{main}, which states that the minimum is found at $x* = (\mu_0+\mu_A)/2$  Then $\xi$ is given by:\\
$\xi = \zeta_\alpha  - \zeta* = \alpha + F_A(x_\alpha) + F_A(x_\alpha) - [1 - F_0(x*) + F_A(x*)]$\\
$= \alpha  + F_A(x_\alpha) + F_0(x*) - F_A(x*)$\\
$= \alpha + [1+ 0.5erf((x_\alpha - x_0)/(\sigma \sqrt{2})] + 0.5erf(\frac{x*-\mu_0}{\sqrt{2}\sigma}) - 0.5erf((x*-\mu_A)/(\sqrt{2}\sigma))$\\
$= \alpha + 1 + 0.5erf((x_\alpha - \mu_0)/(2\sqrt{2}\sigma)) +  0.5erf((\mu_A - \mu_0)/(2\sqrt{2}\sigma)) - 0.5erf((\mu_0 - \mu_A)/(2\sqrt{2}\sigma))$\\,
which can be rewritten in terms of Cohen's d, $d = (\mu_A - \mu_0)/\sigma$, like so:\\

$\xi = \alpha + 1 + \frac{1}{2}erf((x_\alpha - \mu_0)/(2\sqrt{2}\sigma)) +  \frac{1}{2}erf(d/(2\sqrt{2})) - \frac{1}{2}erf(-d/(2\sqrt{2}))$\\

Thus, since the error function $erf$ is known to be monotonically increasing, the last two terms of $\xi$ must be increasing in $d$. Of course, this naturally leaves the question of the term $\frac{1}{2}erf((x_\alpha - \mu_0)/(2\sqrt{2}\sigma))$, which appears to be \textit{decreasing} in $d$, since $x_\alpha$ is a monotonically increasing function of $\mu_0$. However, notice that since we have restricted ourselves to the single-tailed case, $x_\alpha > \mu_0$. So for any value of $\delta$, $\frac{1}{2}erf((x_\alpha - \mu_0)/(2\sqrt{2}\sigma)) < \frac{1}{2}erf((\mu_0 - \mu_A)/(2\sqrt{2}\sigma))$ and the last term sort of ``absorbs" the one that is decreasing in $d$. Hence, for any small increase in $d$, $\xi$ will also increase. Note, in the lower tailed scenario, this would be increasing in $-d$.

\subsection{\normalsize \textbf{F Distribution}}

A natural additional inquiry is that of the performance of our critical value when the test statistic, under the null and alternative hypotheses, follows some kind of an $F$ distribution. This is obviously critical for many results in regression analysis, ANOVA, and any test wherein we use the extra sum of squares principle.\\
Clearly, since the $F$ distribution is not symmetric, it does not satisfy the conditions of Theorem 1, so we cannot use the closed-form solution for the critical value derived therein. Rather, for the purposes of this study, we instead provide specific examples for a few different choices of numerator and denominator degrees of freedom, and non-centrality parameter on the alternative sampling distribution. All of the minimizations were done numerically using the R function optim() \cite{R}.\par

First, we investigate null sampling distribution $f_0 ~ F_{10,10,0}$ and alternative sampling distribution $f_A ~ F_{10,10,10}$. Using the typical method of rejection, that is,  $\alpha = .05$, yields a staggering $\zeta = .78$. That is, if the true non-centrality parameter is $10$ and we have $10$ numerator and denominator degrees of freedom, we have a $78\%$ chance of coming to an incorrect conclusion! Dichotomizing using a critical value that minimizes $\zeta$ reduces this chance to $\zeta = .570$, which is still objectively outrageous, but a dramatic improvement from the above. As can be expected, the probability of a Type I error here is substantially larger than $.05$, at $\alpha = 0.3$. However, the probability of a Type II error is wildly reduced from $\beta_{\alpha = .05} = .732$ to $\beta_{\zeta} = .263$, further illustrating the necessity of a redefined standard for a critical value, in situations where Type I and Type II errors are equally egregious.\par 

Changing the numerator and denominator degrees of freedom to $2$ and $30$, as is more common in practice, makes things better for both methods. The standard method ($\alpha = .05$) yields a Bi Error of $0.278$, and probability of Type II error $\beta_{0.05} = 0.228$. Using a critical value that minimizes the Bi Error reduces the Bi Error to $\zeta = 0.236$, with $P(Type I Error)$ of $\alpha_{\zeta} = 0.112$ and a $P(Type II Error)$ of $ \beta_{\zeta} = 0.123$. Notice in this case as well, even without the subjective error analysis part of the proposed method, critical value $argmin(\zeta)$ is much more appropriate than the classic critical value corresponding to $\alpha = .05$ for situations in which Type I and Type II error are equally egregious.\\

\subsection{\normalsize \textbf{Limitations}}
This method is meant as a replacement for the 5-step method for null hypothesis significance testing. It is meant primarily for students and researchers unfamiliar with statistics; as such its scope is limited with regards to real-world applications. One aspect of this is evident in the fact that Theorm \ref{main} does not apply to F-distributions or the case where the variances of the null and alternative distributions are unequal. Further research is warrented in these areas. 

\section{\Large{Conclusion}}
Misinterpretation of $p$-values and statistical significance testing remains a major problem in modern statistical practice. In order to eradicate the problem at its root, we have defined an accessible replacement for the existing, commonly-known 5-step hypothesis testing procedure. Our method possesses many desirable qualities, including but not limited to simplicity, lack of the need for $p$-values, potential for an inconclusive results outcome, and quantitative incorporation of Type II error. Theoretical and empirical results indicate that this method can be useful in many cases.\\

\section*{Acknowledgements}

We are very grateful to Dr. Lynne Stokes for her feedback on this paper, as well as to the Conference of Texas Statisticians for its support of this work at the poster stage.

\noindent\hrulefill

\renewcommand\refname{REFERENCES}

\appendix
\section{Simulation Results}
In order to assess the performance of each method, we calculated what proportion of the time $H_0$ was rejected, for each value of $n, \mu$, and $\mu_A$. Note, for $\mu = 0$ this represents Type I error, and for $\mu \neq 0$ this proportion represents statistical power. Re-running with a different seed yielded deviances in the third decimal place, so the Monte Carlo error on results is $\pm .01$. Consequently, results are reported to the second decimal place.\par
Vertical columns for the Bi Error method tables represent the hypothesized alternative mean, $\mu_A$. All tables were generated using the R package xtable \cite{xtable}.
\begin{table}[ht]
\centering
\begin{tabular}{rrrrrrr}
\hline
   &     &     & $\mu$&(Actual)&     &     \\ 
  \hline
$\hat{\mu_A} \downarrow $ \vline & 0 & 0.3 & 0.5 & 0.7 & 1 \\ 
  \hline
  0.3 \vline & 0.44 & 0.79 & 0.92 & 0.98 & 1.00 \\ 
  0.5 \vline & 0.40 & 0.76 & 0.91 & 0.98 & 1.00 \\ 
  0.7 \vline & 0.36 & 0.73 & 0.89 & 0.97 & 1.00 \\ 
  1  \vline & 0.32 & 0.67 & 0.86 & 0.96 & 1.00 \\ 
  1.5  \vline & 0.23 & 0.58 & 0.80 & 0.92 & 0.99 \\ 
  2 \vline & 0.17 & 0.49 & 0.73 & 0.89 & 0.98 \\ 
  2.5 \vline & 0.12 & 0.39 & 0.64 & 0.84 & 0.97 \\ 
  3 \vline & 0.08 & 0.32 & 0.55 & 0.76 & 0.95 \\ 
  3.5 \vline & 0.06 & 0.24 & 0.45 & 0.68 & 0.90 \\ 
   \hline
\end{tabular}
\caption{Rejection rates using the Bi Error method, $n=10$, given true  location parameter ($\mu$) and hypothesized alternative location parameter ($\hat{\mu_A}$)}
\end{table}

\begin{table}[H]
\centering
\begin{tabular}{rrrrrrr}
\hline
   &     &     & $\mu$&(Actual)&     &     \\ 
  \hline
 & 0 & 0.3 & 0.5 & 0.7 & 1 \\  
  \hline
& 0.05 & 0.22 & 0.43 & 0.65 & 0.90 \\ 

   \hline
\end{tabular}
\caption{Rejection rates using NHST method with $\alpha = .05$, $n=10$, given true alternative parameter}
\end{table}

\begin{table}[H]
\centering
\begin{tabular}{rrrrrrr}
\hline
   &     &     & $\mu$&(Actual)&     &     \\ 
  \hline
$\hat{\mu_A} \downarrow $ \vline & 0 & 0.3 & 0.5 & 0.7 & 1  \\
  \hline
  0.3 \vline & 0.44 & 0.89 & 0.98 & 1.00 & 1.00 \\ 
  0.5 \vline & 0.40 & 0.86 & 0.98 & 1.00 & 1.00 \\ 
  0.7 \vline & 0.36 & 0.84 & 0.97 & 1.00 & 1.00 \\ 
  1 \vline & 0.31 & 0.81 & 0.96 & 1.00 & 1.00 \\ 
  1.5 \vline & 0.22 & 0.72 & 0.93 & 0.99 & 1.00 \\ 
  2 \vline & 0.17 & 0.63 & 0.89 & 0.98 & 1.00 \\ 
  2.5 \vline & 0.11 & 0.54 & 0.84 & 0.97 & 1.00 \\ 
  3 \vline & 0.07 & 0.45 & 0.77 & 0.94 & 1.00 \\ 
  3.5 \vline & 0.05 & 0.35 & 0.69 & 0.91 & 1.00 \\ 
   \hline
\end{tabular}
\caption{Rejection rates using the Bi Error method, $n=20$, given true  location parameter ($\mu$) and hypothesized alternative location parameter ($\hat{\mu_A}$)}
\end{table}

\begin{table}[H]
\centering
\begin{tabular}{rrrrrrr}
\hline
   &     &     & $\mu$&(Actual)&     &     \\ 
  \hline
& 0 & 0.3 & 0.5 & 0.7 & 1 \\ 
  \hline
& 0.05 & 0.36 & 0.70 & 0.92 & 1.00 \\   

   \hline
\end{tabular}
\caption{Rejection rates using the NHST method with $\alpha = .05$, $n=20$, given true alternative parameter}
\end{table}

\begin{table}[H]
\centering
\begin{tabular}{rrrrrrr}
\hline
   &     &     & $\mu$&(Actual)&     &     \\ 
  \hline
$\hat{\mu_A} \downarrow $ \vline & 0 & 0.3 & 0.5 & 0.7 & 1 \\ 
  \hline
  0.3 \vline & 0.45 & 0.93 & 1.00 & 1.00 & 1.00 \\ 
  0.5 \vline & 0.40 & 0.92 & 0.99 & 1.00 & 1.00 \\ 
  0.7 \vline & 0.37 & 0.90 & 0.99 & 1.00 & 1.00 \\ 
  1 \vline & 0.31 & 0.87 & 0.99 & 1.00 & 1.00 \\ 
  1.5 \vline & 0.23 & 0.82 & 0.98 & 1.00 & 1.00 \\ 
  2 \vline & 0.16 & 0.74 & 0.96 & 1.00 & 1.00 \\ 
  2.5 \vline & 0.11 & 0.66 & 0.93 & 0.99 & 1.00 \\ 
  3 \vline & 0.07 & 0.56 & 0.89 & 0.99 & 1.00 \\ 
  3.5 \vline & 0.04 & 0.46 & 0.84 & 0.98 & 1.00 \\ 
   \hline
\end{tabular}
\caption{Rejection rates using the Bi Error method, $n=30$, given true  location parameter ($\mu$) and hypothesized alternative location parameter ($\hat{\mu_A}$)}
\end{table}

\begin{table}[H]
\centering
\begin{tabular}{rrrrrrr}
\hline
   &     &     & $\mu$&(Actual)&     &     \\ 
  \hline
& 0 & 0.3 & 0.5 & 0.7 & 1 \\ 
  \hline
 & 0.05 & 0.48 & 0.85 & 0.98 & 1.00 \\ 

   \hline
\end{tabular}
\caption{Rejection rates using the NHST method with $\alpha = .05$, $n=30$, given true alternative parameter}
\end{table}

\begin{table}[H]
\centering
\begin{tabular}{rrrrrrr}
\hline
   &     &     & $\mu$&(Actual)&     &     \\ 
  \hline
$\hat{\mu_A} \downarrow $ \vline & 0 & 0.3 & 0.5 & 0.7 & 1 \\ 
  \hline
  0.3 \vline & 0.45 & 0.98 & 1.00 & 1.00 & 1.00 \\ 
  0.5 \vline & 0.41 & 0.97 & 1.00 & 1.00 & 1.00 \\ 
  0.7 \vline & 0.37 & 0.96 & 1.00 & 1.00 & 1.00 \\ 
   1 \vline & 0.32 & 0.95 & 1.00 & 1.00 & 1.00 \\ 
  1.5 \vline & 0.23 & 0.92 & 1.00 & 1.00 & 1.00 \\ 
  2 \vline & 0.17 & 0.87 & 1.00 & 1.00 & 1.00 \\ 
  2.5 \vline & 0.11 & 0.81 & 0.99 & 1.00 & 1.00 \\ 
  3 \vline & 0.07 & 0.74 & 0.98 & 1.00 & 1.00 \\ 
  3.5 \vline & 0.04 & 0.65 & 0.96 & 1.00 & 1.00 \\ 
   \hline
\end{tabular}
\caption{Rejection rates using Bleile's critical value, $n=50$, given true  location parameter ($\mu$) and hypothesized alternative location parameter ($\hat{\mu_A}$)}
\end{table}

\begin{table}[H]
\centering
\begin{tabular}{rrrrrrr}
\hline
   &     &     & $\mu$&(Actual)&     &     \\ 
  \hline
& 0 & 0.3 & 0.5 & 0.7 & 1 \\ 
  \hline
& 0.05 & 0.67 & 0.97 & 1.00 & 1.00 \\ 

   \hline
\end{tabular}
\caption{Rejection rates using the NHST method with $\alpha = .05$, $n=50$, given true alternative parameter}
\end{table}

\onecolumn
\section{Code Appendix}
 The code was run twice, once with the seed apparent here, and once with the seed that is commented out, in order to get an idea of the Monte Carlo error. Displayed tabular results are the aggregation of the two runs.
\begin{lstlisting}[language=R]
compute_CV = function(sample, p1 = .5, alpha = NA,deltaest){
  if(!is.na(alpha)){return(qt(alpha, df=length(sample)-1, ncp = 0, lower.tail = F))}
  else{
    return(deltaest/2 + ((sd(sample))^2*log(p1/(1-p1)))/deltaest)
  }
}

rejection_decision = function(CV, side = "Upper" ,sample, teststat = mean()){
teststat = mean(sample)*sqrt(n)/sd(sample)
  if(side == "Upper"){if(teststat>CV){return(1)}
                        else{return(0)}}
  else if(side == "Lower"){if(teststat< CV){return(1)}else{return(0)}}
  else if (side == "Two tailed"){if(abs(teststat)> CV){return(1)}else{return(0)}}
  else{return("Invalid Input")}
}

make_table = function(tab, arr,deltas,mus,ns,n,M){
  for(d in deltas){
    for(m in mus){
      i = which(deltas==d)
      j = which(mus ==m)
      k= which(ns ==n)
      tab[i,j] = sum(arr[,k,j,i])/M
      rownames(tab) = deltas
      colnames(tab) = mus
    }
  }
  return(tab)
}

aggregate_table = function(tab){
  newtab = matrix(nrow = 1, ncol = ncol(tab))
  colnames(newtab) = colnames(tab)
  for(i in 1:ncol(tab)){
    newtab[1,i] = mean(tab[1:nrow(tab),i])
  }
  return(newtab)
}

source("functions.R")
  set.seed(198663)
  #set.seed(467732)
  M = 10000
  ns = c(10,20,30,50)
  mus = c(0,0.3,0.5,0.7,1)
  deltas = c(0.1,0.3,0.5,0.7,1,1.5,2,2.5,3,3.5)
  MLmethod_out = array(dim= list(M, length(ns), length(mus), length(deltas)), 
              dimnames = list(1:M, ns, mus, deltas))
    
  StandardMethod_out = array(dim= list(M, length(ns), length(mus), length(deltas)), 
                             dimnames = list(1:M, ns, mus,deltas))
  for(i in 1:M){
    for(d in deltas){
    for(n in ns){
      for(mu in mus){
        samp=rnorm(n, mu)
        CVml = compute_CV(samp, deltaest=d)
        CVa = compute_CV(samp, alpha = .05, deltaest = d)
        MLmethod_out[i,which(ns==n), which(mus==mu), which(deltas==d)] =
        rejection_decision(CVml,sample=samp)
        
        StandardMethod_out[i, which(ns==n),which(mus==mu),which(deltas==d)] =
        rejection_decision(CVa,sample = samp)
      }
      }
    }
  }
  \end{lstlisting}

\end{document}